\documentclass[12pt, reqno]{amsart}
\usepackage{amsmath, amsthm, amscd, amsfonts, amssymb, graphicx, color}
\usepackage[bookmarksnumbered, colorlinks, plainpages]{hyperref}

\hypersetup{colorlinks=true,linkcolor=red, anchorcolor=green, citecolor=cyan, urlcolor=red, filecolor=magenta, pdftoolbar=true}
\theoremstyle{definition}

\theoremstyle{remark}

\numberwithin{equation}{section}
\input{tcilatex}

\begin{document}
\title[\textbf{A Profit-maximization Model for a Company}]{\textbf{A
Profit-maximization Model for a Company that Sells an Arbitrary Number of
Products }}
\author[D.-P. Covei \textit{and I. Gheorghe-Iv\u{a}nescu}]{\textit{Drago\c{s}%
-P\u{a}tru Covei}$^{1}$\textit{\ and Ioan Gheorghe-Iv\u{a}nescu}$^{2}$%
\textsf{\ }}
\address{ $^{1}${\small \textit{Department of Applied Mathematics}}\\
{\small \ \textit{The Bucharest University of Economic Studies }}\\
{\small \textit{Piata Romana, 1st district, postal code: 010374, postal
office: 22, Romania}}}
\email{\textcolor[rgb]{0.00,0.00,0.84}{coveid@yahoo.com}}
\address{$^{2}${\small \textit{The Bucharest University of Economic Studies }%
}\\
{\small \textit{Piata Romana, 1st district, postal code: 010374, postal
office: 22, Romania}}}
\date{Received: xxxxxx; Revised: yyyyyy; Accepted: zzzzzz. \\
\indent $^{*}$Corresponding author}

\begin{abstract}
One of the problems faced by a firm that sells certain commodities is to
determine the number of products that it must supply in order to maximize
its profit. In this article, the authors give an answer to this problem of
economic interest. The proposed problem is a generalization of the results
obtained by Stirzaker (Probability and Random Variables: A Beginner's Guide,
1999) and Kupferman (Lecture Notes in Probability, 2009) where the authors
do not present a situation where the sale of a quantity from some
commodities is constrained by the marketing of another. In addition, the
described procedure is simple and can be successfully applied to any number
of commodities. The obtained results can be easily put into practice.
\end{abstract}

\maketitle

\setcounter{page}{1}


\let\thefootnote\relax\footnote{%
Copyright 2016 by the JOURNAL.}


\section{Introduction}

Maximization problems arise in various fields of science, either in a direct
form or in an indirect one. The objective of this paper is to offer an
answer to the problem of maximizing the profit of a company that sells
certain products in a competitive market under the assumption that previous
statistics conducted by the firm, establish, as accurately as possible, the
probability of purchasing products. More precisely, we will prove the
following results:

\textbf{Theorem 1.} \textit{Let us suppose a company wants to supply with
two commodities: }$M_{i}$\textit{, }$i=1,2$\textit{, whose sale on the
market brings the company a profit of }$c_{i}$\textit{\ Euros/product if the
product sells and a loss of }$s_{i}$\textit{\ Euros/product if the product
does not sell. Moreover, let }$X$\textit{\ be a continuous random variable
with the density function }$f_{X}\left( x\right) $\textit{\ that summarises
the demand for commodity }$M_{1}$\textit{\ and distribution function }$%
F_{X}\left( x\right) $\textit{\ and }$Y$\textit{\ a continuous random
variable with the density function }$f_{Y}\left( y\right) $\textit{\ that
summarises the demand for commodity }$M_{2}$\textit{\ with the distribution
function }$F_{Y}\left( y\right) $\textit{\ and }$\left( X,Y\right) $\textit{%
\ a continuous random vector with the density function }$f_{X,Y}\left(
x,y\right) $\textit{. If }$n_{i}$\textit{\ (}$i=1,2$\textit{) is the number
of products from the commodity }$M_{i}$\textit{\ that are about to be
ordered by the company then the company maximizes its profit when }$n_{1}$%
\textit{\ is the point where the distribution function }$F_{X}\left(
x\right) $\textit{\ reaches the level }$c_{1}/\left( c_{1}+s_{1}\right) $%
\textit{\ the way }$x$\textit{\ grows and }$n_{2}$\textit{\ is the point
where the distribution function }$F_{Y}\left( y\right) $\textit{\ reaches
the level }$c_{2}/\left( c_{2}+s_{2}\right) $\textit{\ the way }$y$\textit{\
grows.}

We note that another problem facing us is to choose the number of products,
given the profit function and, given the "constraints" imposed by certain
factors that may occur. We formulate, thus, the following problem:

\textbf{Problem 1}.\textit{\ If there are restrictions for }$n_{1}$\textit{\
and }$n_{2}$\textit{\ such as }$f\left( n_{1}\text{,}n_{2}\right) \leq 0$%
\textit{\ let us write the maximization problem in order to determine the
quantity }$\left( n_{1},n_{2}\right) $\textit{\ from the commodities }$%
\left( M_{1}\text{,}M_{2}\right) $\textit{\ you have to order so that the
company will maximize the profit and to suggest a method to solve the above
formulated problem.}

Note that the ineguality $f\left( n_{1}\text{,}n_{2}\right) \leq 0$
summarises constraints of the number of products due to the problems faced
by the firm.

Since in Theorem 1 the application is represented if the random variables
are continuous, it is natural to ask what happens if they are discrete.

The answer is given in the following theorem:

\textbf{Theorem 2. }\textit{Let us suppose that a company wants to supply
with two commodities: }$M_{i}$\textit{, }$i=1,2$\textit{, whose marketing
brings a profit of }$c_{i}$\textit{\ Euros/product if the product sells and
a loss of }$s_{i}$\textit{\ Euros/product if the product does not sell. Let }%
$X$\textit{\ be a discrete random variable with probability mass function }$%
p_{X}\left( x\right) $\textit{\ that summarises the demand for commodity }$%
M_{1}$\textit{, }$Y$\textit{, a discrete random variable with probability
mass function }$p_{Y}\left( y\right) $\textit{\ that summarises the demand
for commodity }$M_{2}$\textit{\ and }$\left( X,Y\right) $\textit{\ a
discrete random vector with probability common mass function }$p\left(
i,j\right) =P\left( X=i,Y=j\right) $\textit{, }$i,j=0,1,...$

\textit{If }$n_{i}$\textit{\ (}$i=1,2$\textit{) is the number of products
from the commodity }$M_{i}$\textit{\ that are about to be ordered by the
company then:}

\textit{i)\quad If the company net profit resulted from the marketing of the
two commodity }$M_{1}$\textit{, }$M_{2}$\textit{\ is given by the real
function }$g_{X,Y}\left( n_{1},n_{2}\right) $\textit{\ then the following
equality takes place}%
\begin{equation*}
M\left[ g_{X,Y}\left( n_{1},n_{2}\right) \right] =M\left[ g_{1}\left(
n_{1},X\right) \right] +M\left[ g_{2}\left( n_{2},Y\right) \right] 
\end{equation*}%
\textit{where }$g_{1}\left( n_{1},X\right) $\textit{\ is the net profit
brought to the company by the commodity }$M_{1}$\textit{\ and }$g_{2}\left(
n_{2},Y\right) $\textit{\ is the net profit brought to the company by the
commodities }$M_{2}$\textit{.}

\textit{ii)\quad As long as }%
\begin{equation*}
\left( c_{1}+s_{1}\right) P\left( X\leq n_{1}\right) <c_{1}\text{ and }%
\left( c_{2}+s_{2}\right) P\left( Y\leq n_{2}\right) <c_{2}
\end{equation*}%
\textit{the company will maximize the profit.}

Let us notice that, in the situation in which there are restrictions imposed
on $n_{1}$ and $n_{2}$ like $f\left( n_{1}\text{,}n_{2}\right) =0$ then the
maximum number of products $\left( n_{1},n_{2}\right) $, from the
commodities $M_{1}$ and $M_{2}$ that must be ordered in order for the
company to maximize the profit, is determined by

\begin{equation}
\underset{n_{1},n_{2}\geq 0}{\max }M\left[ g_{X,Y}\left( n_{1},n_{2}\right) %
\right] \text{ with constraints }f\left( n_{1}\text{,}n_{2}\right) =0.
\label{maxl2}
\end{equation}%
\textbf{Theorem 2} leads to the following problem:

\textbf{Problem 2.} \textit{Can we give an answer at the maximization
problem (\ref{maxl2})?}

The questions one and two appear, for example, inside the companies that
produce both systems for video games and video games. For instance, the
games can be compatible with older systems, which means that even if the
company does not sell new systems, it will still have demands for the new
games. Thus, the constraints of production, supply and transport for the
systems imply the fact that for a large number of sold systems there is a
light decrease in the amount of profit per each sold game and therefore,
taking into consideration the company interest to maximize the profit, the
relations like $f\left( n_{1},n_{2}\right) \leq 0$ are a must.

\textbf{Remark 1}: \textit{The optimization problem we consider in \textbf{%
Theorem 1} and \textbf{Theorem 2} is to maximize the expected profit from
selling some commodities. Other extensions of this work encompasses
optimization criterions which involve higher moments of profit such as
variance, skewness, kurtosis. For instance an interesting problem is to
minimize the variance of the profit. We leave these extensions for future
research. }

Let us note that the results from Theorems 1 and 2 are also found in a
particular way in Stirzaker [2] and Kupferman [3].

We are in the position to give an answer at the mentioned theorems.

\section{Proof of Theorem 1}

i) The net profit brought to the company by the two commodities is given by%
\begin{equation*}
g_{X,Y}\left( n_{1},n_{2}\right) =\left\{ 
\begin{array}{l}
Xc_{1}-\left( n_{1}-X\right) s_{1}\text{ }+Yc_{2}-\left( n_{2}-Y\right) s_{2}%
\text{ if }X\leq n_{1}\text{, }Y\leq n_{2} \\ 
Xc_{1}-\left( n_{1}-X\right) s_{1}\text{ }+c_{2}n_{2}\text{ if }X\leq n_{1}%
\text{ and }Y>n_{2} \\ 
c_{1}n_{1}\text{ }+Yc_{2}-\left( n_{2}-Y\right) s_{2}\text{ if }X>n_{1}\text{
and }Y\leq n_{2} \\ 
c_{1}n_{1}+c_{2}n_{2}\text{ if }X>n_{1}\text{ and }Y>n_{2}.%
\end{array}%
\right. 
\end{equation*}%
Or, written in a different way%
\begin{equation}
g_{X,Y}\left( n_{1},n_{2}\right) =\left\{ 
\begin{array}{l}
X\left( c_{1}+s_{1}\text{ }\right) +Y\left( c_{2}+s_{2}\right)
-n_{1}s_{1}-n_{2}s_{2}\text{ if }X\leq n_{1}\text{, }Y\leq n_{2} \\ 
X\left( c_{1}+s_{1}\right) -n_{1}s_{1}\text{ }+c_{2}n_{2}\text{ if }X\leq
n_{1}\text{ and }Y>n_{2} \\ 
Y\left( c_{2}+s_{2}\right) +c_{1}n_{1}\text{ }-n_{2}s_{2}\text{ if }X>n_{1}%
\text{ and }Y\leq n_{2} \\ 
c_{1}n_{1}+c_{2}n_{2}\text{ if }X>n_{1}\text{ and }Y>n_{2}\text{.}%
\end{array}%
\right.   \label{profn}
\end{equation}%
The expected gain is deduced using the law of the unconscious statistician,
by calculating 
\begin{eqnarray*}
M\left[ g_{X,Y}\left( n_{1},n_{2}\right) \right]  &=&\int_{-\infty
}^{n_{1}}\int_{-\infty }^{n_{2}}\left[ x\left( c_{1}+s_{1}\text{ }\right)
+y\left( c_{2}+s_{2}\right) -n_{1}s_{1}-n_{2}s_{2}\right] f_{X,Y}\left(
x,y\right) dxdy \\
&&+\int_{-\infty }^{n_{1}}\int_{n_{2}}^{\infty }\left[ x\left(
c_{1}+s_{1}\right) -n_{1}s_{1}\text{ }+c_{2}n_{2}\text{ }\right]
f_{X,Y}\left( x,y\right) dydx \\
&&+\int_{n_{1}}^{\infty }\int_{-\infty }^{n_{2}}\left[ y\left(
c_{2}+s_{2}\right) +c_{1}n_{1}\text{ }-n_{2}s_{2}\text{ }\right]
f_{X,Y}\left( x,y\right) dydx \\
&&+\int_{n_{1}}^{\infty }\int_{n_{2}}^{\infty }\left( c_{1}n_{1}+c_{2}n_{2}%
\text{ }\right) f_{X,Y}\left( x,y\right) dxdy
\end{eqnarray*}%
or, echivalently%
\begin{eqnarray*}
M\left[ g_{X,Y}\left( n_{1},n_{2}\right) \right]  &=&\int_{-\infty }^{n_{1}}%
\left[ xc_{1}-\left( n_{1}-x\right) s_{1}\right] f_{X}\left( x\right)
dx+\int_{n_{1}}^{\infty }n_{1}c_{1}f_{X}\left( x\right) dx \\
&&+\int_{-\infty }^{n_{2}}\left[ yc_{2}-\left( n_{2}-y\right) s_{2}\right]
f_{Y}\left( y\right) dy+\int_{n_{2}}^{\infty }n_{2}c_{2}f_{Y}\left( y\right)
dx \\
&=&c_{1}n_{1}+\left( c_{1}+s_{1}\right) \int_{0}^{n_{1}}\left(
x-n_{1}\right) f_{X}\left( x\right) dx \\
&&+c_{2}n_{2}+\left( c_{2}+s_{2}\right) \int_{0}^{n_{2}}\left(
y-n_{2}\right) f_{Y}\left( y\right) dx.
\end{eqnarray*}%
We need to maximize this expression with respect to $n_{1}$ and $n_{2}$. The
simplest way to do it is find the critical point for this expression $M\left[
g_{X,Y}\left( n_{1},n_{2}\right) \right] $, thus 
\begin{equation*}
\left\{ 
\begin{array}{c}
M_{n_{1}}^{\prime }\left[ g_{X,Y}\left( n_{1},n_{2}\right) \right] =0 \\ 
M_{n_{2}}^{\prime }\left[ g_{X,Y}\left( n_{1},n_{2}\right) \right] =0%
\end{array}%
\right. 
\end{equation*}%
equivalently%
\begin{equation*}
\left\{ 
\begin{array}{c}
c_{1}+\left( c_{1}+s_{1}\right) \int_{0}^{n_{1}}f_{X}\left( x\right)
dx=c_{1}-\left( c_{1}+s_{1}\right) F_{X}\left( n_{1}\right) =0 \\ 
c_{2}+\left( c_{2}+s_{2}\right) \int_{0}^{n_{2}}f_{Y}\left( y\right)
dy=c_{2}-\left( c_{2}+s_{2}\right) F_{Y}\left( n_{2}\right) =0%
\end{array}%
\right. 
\end{equation*}%
from which we get that the critical point verifies the following%
\begin{equation}
F_{X}\left( n_{1}\right) =c_{1}/\left( c_{1}+s_{1}\right) \text{ and }%
F_{Y}\left( n_{2}\right) =c_{2}/\left( c_{2}+s_{2}\right) \text{.}
\label{crit}
\end{equation}%
In order to establish if the determined critical point is maximum, we write
the hessian matrix%
\begin{equation*}
H_{M}\left( n_{1},n_{2}\right) =\left( 
\begin{array}{cc}
-\left( c_{1}+s_{1}\right) f_{X}\left( n_{1}\right)  & 0 \\ 
0 & -\left( c_{2}+s_{2}\right) f_{Y}\left( n_{2}\right) 
\end{array}%
\right) .
\end{equation*}%
We observe that $\Delta _{1}=-\left( c_{1}+s_{1}\right) f_{X}\left(
n_{1}\right) <0$ and%
\begin{eqnarray*}
\Delta _{2} &=&\left\vert 
\begin{array}{cc}
-\left( c_{1}+s_{1}\right) f_{X}\left( n_{1}\right)  & 0 \\ 
0 & -\left( c_{2}+s_{2}\right) f_{Y}\left( n_{2}\right) 
\end{array}%
\right\vert  \\
&=&\left( c_{1}+s_{1}\right) \left( c_{2}+s_{2}\right) f_{X}\left(
n_{1}\right) f_{Y}\left( n_{2}\right) >0.
\end{eqnarray*}%
From which we deduce that the function $M\left[ g_{X,Y}\left(
n_{1},n_{2}\right) \right] $ is strictly concave, meaning that the critical
point obtained from (\ref{crit}) is the global maximum point for which the
company should order $n_{1}$ products from the commodities $M_{1}$ and $n_{2}
$ products from the commodities $M_{2}$, where $n_{1}$ is the point in which
the distribution function $F_{X}\left( x\right) $ reaches the level $%
c_{1}/\left( c_{1}+s_{1}\right) $ the $x$\ way grows and $n_{2}$ is the
point in which the distribution function $F_{Y}\left( y\right) $ reaches the
level $c_{2}/\left( c_{2}+s_{2}\right) $ the $y$ way grows.

Question 1 boils down to proving 
\begin{equation*}
\underset{n_{1},n_{2}\geq 0}{\max }M\left[ g_{X,Y}\left( n_{1},n_{2}\right) %
\right] \text{ with constraint }f\left( n_{1}\text{,}n_{2}\right) \leq 0.
\end{equation*}%
That can be solved, for example, using Lagrange multipliers method if $%
f\left( n_{1},n_{2}\right) =0$ or in the situation $f\left( n_{1}\text{,}%
n_{2}\right) <0$ using other methods that can be found in reference of
Intriligator [1].

\section{Proof of Theorem 2}

i) The net profit of the company is given by the function $g_{X,Y}\left(
n_{1},n_{2}\right) $ defined through 
\begin{equation}
g_{X,Y}\left( n_{1},n_{2}\right) =\left\{ 
\begin{array}{l}
X\left( c_{1}+s_{1}\text{ }\right) +Y\left( c_{2}+s_{2}\right)
-n_{1}s_{1}-n_{2}s_{2}\text{ if }X=\overline{0,n_{1}}\text{, }Y=\overline{%
0,n_{2}} \\ 
X\left( c_{1}+s_{1}\right) -n_{1}s_{1}\text{ }+c_{2}n_{2}\text{ if }%
X=0,...,n_{1}\text{ and }Y>n_{2}+1 \\ 
Y\left( c_{2}+s_{2}\right) +c_{1}n_{1}\text{ }-n_{2}s_{2}\text{ if }X>n_{1}%
\text{ and }Y=0,...,n_{2} \\ 
c_{1}n_{1}+c_{2}n_{2}\text{ if }X>n_{1}\text{ and }Y>n_{2}\text{.}%
\end{array}%
\right.   \label{profnd}
\end{equation}%
Using \textbf{The extended law of the unconscious statistician }we deduce
that the realised profit mean value by the company through the two products
marketing is%
\begin{equation*}
\begin{array}{ll}
M\left[ g_{X,Y}\left( n_{1},n_{2}\right) \right] = & \overset{n_{1}}{%
\underset{i=0}{\sum }}\overset{n_{2}}{\underset{j=0}{\sum }}\left[ i\left(
c_{1}+s_{1}\text{ }\right) +j\left( c_{2}+s_{2}\right) -n_{1}s_{1}-n_{2}s_{2}%
\right] p\left( i,j\right)  \\ 
& +\overset{n_{1}}{\underset{i=0}{\sum }}\overset{\infty }{\underset{%
j=n_{2}+1}{\sum }}\left[ i\left( c_{1}+s_{1}\right) -n_{1}s_{1}\text{ }%
+c_{2}n_{2}\right] p\left( i,j\right)  \\ 
& +\overset{\infty }{\underset{i=n_{1}+1}{\sum }}\overset{n_{2}}{\underset{%
j=0}{\sum }}\left[ j\left( c_{2}+s_{2}\right) +c_{1}n_{1}\text{ }-n_{2}s_{2}%
\right] p\left( i,j\right)  \\ 
& +\overset{\infty }{\underset{i=n_{1}+1}{\sum }}\overset{\infty }{\underset{%
j=n_{2}+1}{\sum }}\left( c_{1}n_{1}+c_{2}n_{2}\text{ }\right) p\left(
i,j\right)  \\ 
& =\left( c_{1}+s_{1}\text{ }\right) \overset{n_{1}}{\underset{i=0}{\sum }}%
\overset{n_{2}}{\underset{j=0}{\sum }}ip\left( i,j\right) +\left( c_{2}+s_{2}%
\text{ }\right) \overset{n_{1}}{\underset{i=0}{\Sigma }}\overset{n_{2}}{%
\underset{j=0}{\Sigma }}jp\left( i,j\right)  \\ 
& -\left( n_{1}s_{1}+n_{2}s_{2}\right) \overset{n_{1}}{\underset{i=0}{\sum }}%
\overset{n_{2}}{\underset{j=0}{\sum }}p\left( i,j\right)  \\ 
& +\left( c_{1}+s_{1}\right) \overset{n_{1}}{\underset{i=0}{\sum }}\overset{%
\infty }{\underset{j=n_{2}+1}{\sum }}ip\left( i,j\right) +\left( -n_{1}s_{1}%
\text{ }+c_{2}n_{2}\right) \overset{n_{1}}{\underset{i=0}{\sum }}\overset{%
\infty }{\underset{j=n_{2}+1}{\sum }}p\left( i,j\right)  \\ 
& +\left( c_{2}+s_{2}\right) \overset{\infty }{\underset{i=n_{1}+1}{\sum }}%
\overset{n_{2}}{\underset{j=0}{\sum }}jp\left( i,j\right) +\left( c_{1}n_{1}%
\text{ }-n_{2}s_{2}\right) \overset{\infty }{\underset{i=n_{1}+1}{\sum }}%
\overset{n_{2}}{\underset{j=0}{\sum }}p\left( i,j\right)  \\ 
& +\left( c_{1}n_{1}+c_{2}n_{2}\text{ }\right) \overset{\infty }{\underset{%
i=n_{1}+1}{\sum }}\overset{\infty }{\underset{j=n_{2}+1}{\sum }}p\left(
i,j\right) 
\end{array}%
\end{equation*}%
or, written as%
\begin{equation*}
\begin{array}{c}
M\left[ g_{X,Y}\left( n_{1},n_{2}\right) \right] =c_{1}n_{1}\overset{\infty }%
{\underset{i=n_{1}+1}{\sum }}\overset{\infty }{\underset{j=0}{\sum }}p\left(
i,j\right) +c_{2}n_{2}\text{ }\overset{\infty }{\underset{i=0}{\sum }}%
\overset{\infty }{\underset{j=n_{2}+1}{\sum }}p\left( i,j\right)  \\ 
-n_{1}s_{1}\text{ }\overset{n_{1}}{\underset{i=0}{\sum }}\overset{\infty }{%
\underset{j=0}{\sum }}p\left( i,j\right) -n_{2}s_{2}\overset{\infty }{%
\underset{i=0}{\sum }}\overset{n_{2}}{\underset{j=0}{\sum }}p\left(
i,j\right)  \\ 
+\left( c_{2}+s_{2}\right) \overset{\infty }{\underset{i=0}{\sum }}\overset{%
n_{2}}{\underset{j=0}{\sum }}jp\left( i,j\right) \text{ }+\left(
c_{1}+s_{1}\right) \overset{n_{1}}{\underset{i=0}{\sum }}\overset{\infty }{%
\underset{j=0}{\sum }}ip\left( i,j\right) 
\end{array}%
\end{equation*}%
from which%
\begin{equation*}
\begin{array}{c}
M\left[ g_{X,Y}\left( n_{1},n_{2}\right) \right] =c_{1}n_{1}\overset{\infty }%
{\underset{i=n_{1}+1}{\sum }}p_{X}\left( i\right) +c_{2}n_{2}\overset{\infty 
}{\underset{j=n_{2}+1}{\sum }}p_{Y}\left( j\right)  \\ 
-n_{1}s_{1}\text{ }\overset{n_{1}}{\underset{i=0}{\sum }}p_{X}\left(
i\right) -n_{2}s_{2}\overset{n_{2}}{\underset{j=0}{\sum }}p_{Y}\left(
j\right) +\left( c_{2}+s_{2}\right) \overset{n_{2}}{\underset{j=0}{\sum }}%
jp_{Y}\left( j\right) +\left( c_{1}+s_{1}\right) \overset{n_{1}}{\underset{%
i=0}{\sum }}ip_{X}\left( i\right) 
\end{array}%
\end{equation*}%
and finally 
\begin{eqnarray*}
M\left[ g_{X,Y}\left( n_{1},n_{2}\right) \right]  &=&-n_{1}s_{1}\text{ }%
\overset{n_{1}}{\underset{i=0}{\sum }}p_{X}\left( i\right) +c_{1}n_{1}%
\overset{\infty }{\underset{i=n_{1}+1}{\sum }}p_{X}\left( i\right) +\left(
c_{1}+s_{1}\right) \overset{n_{1}}{\underset{i=0}{\sum }}ip_{X}\left(
i\right)  \\
&&-n_{2}s_{2}\overset{n_{2}}{\underset{j=0}{\sum }}p_{Y}\left( j\right)
+c_{2}n_{2}\overset{\infty }{\underset{j=n_{2}+1}{\sum }}p_{Y}\left(
j\right) +\left( c_{2}+s_{2}\right) \overset{n_{2}}{\underset{j=0}{\sum }}%
jp_{Y}\left( j\right)  \\
&=&c_{1}n_{1}\overset{\infty }{\underset{i=0}{\sum }}p_{X}\left( i\right)
+\left( c_{1}+s_{1}\right) \overset{n_{1}}{\underset{i=0}{\sum }}\left(
i-n_{1}\right) p_{X}\left( i\right)  \\
&&+c_{2}n_{2}\overset{\infty }{\underset{j=0}{\sum }}p_{Y}\left( j\right)
+\left( c_{2}+s_{2}\right) \overset{n_{2}}{\underset{j=0}{\sum }}\left(
j-n_{2}\right) p_{Y}\left( j\right)  \\
&=&c_{1}n_{1}+c_{2}n_{2}+\left( c_{1}+s_{1}\right) \overset{n_{1}}{\underset{%
i=0}{\sum }}\left( i-n_{1}\right) p_{X}\left( i\right) +\left(
c_{2}+s_{2}\right) \overset{n_{2}}{\underset{j=0}{\sum }}\left(
j-n_{2}\right) p_{Y}\left( j\right) .
\end{eqnarray*}%
On the other hand, using [3] we can see that%
\begin{equation*}
M\left[ g_{1}\left( n_{1},X\right) \right] =c_{1}n_{1}+\left(
c_{1}+s_{1}\right) \overset{n_{1}}{\underset{i=0}{\sum }}\left(
i-n_{1}\right) p_{X}\left( i\right) 
\end{equation*}%
and%
\begin{equation*}
M\left[ g_{2}\left( n_{2},Y\right) \right] =c_{2}n_{2}+\left(
c_{2}+s_{2}\right) \overset{n_{2}}{\underset{j=0}{\sum }}\left(
j-n_{2}\right) p_{Y}\left( j\right) 
\end{equation*}%
from which we have $M\left[ g_{X,Y}\left( n_{1},n_{2}\right) \right] =M\left[
g_{1}\left( n_{1},X\right) \right] +M\left[ g_{2}\left( n_{2},Y\right) %
\right] $, which confirm our intuition.

ii) Setting%
\begin{equation*}
G\left( n_{1},n_{2}\right) :=c_{1}n_{1}+\left( c_{1}+s_{1}\right) \overset{%
n_{1}}{\underset{i=0}{\sum }}\left( i-n_{1}\right) p_{X}\left( i\right)
+c_{2}n_{2}+\left( c_{2}+s_{2}\right) \overset{n_{2}}{\underset{j=0}{\sum }}%
\left( j-n_{2}\right) p_{Y}\left( j\right) 
\end{equation*}%
we observe that 
\begin{equation*}
G\left( n_{1}+1,n_{2}+1\right) -G\left( n_{1},n_{2}\right) =c_{1}-\left(
c_{1}+s_{1}\right) \overset{n_{1}}{\underset{i=0}{\sum }}p_{X}\left(
i\right) +c_{2}-\left( c_{2}+s_{2}\right) \overset{n_{2}}{\underset{j=0}{%
\sum }}p_{Y}\left( j\right) .
\end{equation*}%
On the other hand, if%
\begin{equation}
P\left( X\leq n_{1}\right) =\overset{n_{1}}{\underset{i=0}{\sum }}%
p_{X}\left( i\right) <\frac{c_{1}}{c_{1}+s_{1}}\text{ and }P\left( Y\leq
n_{2}\right) =\overset{n_{2}}{\underset{j=0}{\sum }}p_{Y}\left( j\right) <%
\frac{c_{2}}{c_{2}+s_{2}}  \label{maxp}
\end{equation}%
it is obvious that%
\begin{equation*}
G\left( n_{1}+1,n_{2}+1\right) -G\left( n_{1},n_{2}\right) =M\left[
g_{X,Y}\left( n_{1}+1,n_{2}+1\right) \right] -M\left[ g_{X,Y}\left(
n_{1},n_{2}\right) \right] >0.
\end{equation*}%
Absolutely analog%
\begin{equation*}
G\left( n_{1}+1,n_{2}\right) -G\left( n_{1},n_{2}\right) >0\text{ and }%
G\left( n_{1},n_{2}+1\right) -G\left( n_{1},n_{2}\right) >0.
\end{equation*}%
Furthermore, the pair $\left( n_{1},n_{2}\right) $ with $n_{1}$, $n_{2}$ the
largest possible that verifies (\ref{maxp}) achieving the maximum value of $M%
\left[ g_{X,Y}\left( n_{1},n_{2}\right) \right] $ and thus the profit
maximization of the company.

As far as Problem 2 is concerned, to solve (\ref{maxl2}) the analysis splits
into two cases:

Case 1: If one of the variables can be written explicitly as a function of
the other, then by substituting it in the profit function leads to an
unconstrained optimization.

Case 2: If Case 1 does not hold then we can follow the methods presented in
Intriligator [1].

\end{document}